\documentstyle[aps,epsfig,amsfonts]{revtex}

\font\grbit=cmmib10 scaled1000
\newcommand{\gbi}[1]{\mbox{\grbit\symbol{#1}}}
\def\omegabi{{\gbi{'041}}}
\def\comment#1{}

\newcommand{\tea}{time evolution amplitude}

\newcommand{\mn}[1]{\marginpar{\tiny#1}}

\def\cm#1{}

\everymath={\displaystyle}

\newcommand{\p}{\partial}          
\newcommand{\f}[2]{\frac{#1}{#2}}
\newcommand{\be}{\begin{eqnarray}}
\newcommand{\ee}{\end{eqnarray}}
\newcommand{\beqn}{\begin{eqnarray}}
\newcommand{\eeqn}{\end{eqnarray}}
\newcommand{\la}{\langle}
\newcommand{\ra}{\rangle}
\newcommand{\xa}{x_a}
\newcommand{\xaplus}{x_{a^+}}
\newcommand{\xb}{x_b}
\newcommand{\xc}{x}
\newcommand{\ta}{t_a}
\newcommand{\taplus}{t_{a^+}}
\newcommand{\tb}{t_b}
\newcommand{\tc}{t}
\newcommand{\hatU}{\hat{U}}
\newcommand{\calD}{{\cal D}}
\newcommand{\calA}{{\cal A}}
\newcommand{\calAcl}{{\rm A}}
\newcommand{\intab}{\int_{\ta}^{\tb}}
\newcommand{\fluc}{{F}}
\newcommand{\xs}{\tilde{x}}
\newcommand{\xcl}{{x}_{\rm cl}}
\newcommand{\xcs}{\tilde{x}}
\newcommand{\dx}{\delta x}
\newcommand{\dotx}{\dot{x}}
\newcommand{\doty}{\dot{y}}
\newcommand{\dotB}{\dot{B}}
\newcommand{\one}{1\!\!1}

\title{Global Derivation of the Fluctuation Determinant\\ from Group
  Property of Time Evolution.}

\author{H. Kleinert \and B. Van den Bossche\thanks{Alexander von Humboldt
 Fellow}\thanks{Collaborateur scientifique
 du FNRS, On leave from absence of
 Physique Nucl\'eaire Th\'eorique, B5, Universit\'e de Li\`ege Sart-Tilman,
 4000 Li\`ege, Belgium}}

\address{Institut f\"ur Theoretische Physik, Arnimallee 14 D-14195 Berlin,
 Germany}

\begin{document}

\maketitle

\begin{abstract}
The  Van Vleck-Pauli-Morette
fluctuation determinant
is derived
from the global
group property
of the
time evolution amplitude
in a continuous formulation of path integrals.
\end{abstract}

\section{Introduction}
In the semiclassical limit, the
quantum mechanical time evolution amplitude
consists of an exponential of the classical action
$\exp{(i\calAcl)}$, multiplied by a fluctuation factor
$\fluc$
containing the inverse square root of the functional
fluctuation determinant of harmonic eigenmodes of the system.
The standard derivations of $F$ rely on the
local group
properties of the time evolution amplitude \cite{pi}. The initial historic
paper of DeWitt-Morette
\cite{vvpm}
determined $F$ by enforcing these properties for infinitesimal time slices
of the amplitude,
which are necessarily
 semiclassical
by Dirac's observation
\cite{dirac}.
The full fluctuation factor $F$ for finite times
was then composed by a  limiting procedure
from the
$F$'s of the
time slices.
The result was expressed as a square root of an ordinary
matrix determinant, the
 Van Vleck-Pauli-Morette determinant \cite{vvpm}.

Later, Gelfand and Yaglom \cite{gy} related the fluctuation determinant
to the  solution of a second-order differential
equation, again via time-slicing techniques.
This solution  can, of course, be related
to
the
 Van Vleck-Pauli-Morette determinant.

In this note, we point out a more compact
 way of obtaining the fluctuation factor
from
the global, finite-time
group property of
the  time evolution operator.
Only the  continuum
formulation
of path integrals is used.
Our derivation involves neither
time-sliced actions nor differential equations.

\section{Semiclassical approximation}

In Schr\"odinger theory,
a point particle
in a $D$-dimensional euclidean space
 has associated with it a
 Hamilton operator $\hat H(t)$, and a
time evolution operator
 $\hatU(t)$,
 which
determines the
 amplitude to go from a position
$\xa$
at time $\ta$ to
a position $\xb$
at time $\tb$ by the matrix elements~\cite{pi}
\be
( \xb\tb|\xa\ta )=\theta(\tb-\ta)\la\xb|\hatU(\tb,\ta)|\xa\ra.
\label{eq0}
\ee
The Heaviside function $\theta(\tb-\ta)$ ensures causality by
 the
 vanishing of the
amplitude for times $\tb<\ta$.
As elements of a one-parameter Lie group,
the time evolution operators for different times
satisfy the
group multiplication law
\be
\hatU(\tb,\ta)=\hatU(\tb,\tc)\hatU(\tc,\ta).
\label{eq00}
\ee
For matrix elements, this reads
\beqn
\la\xb|\hatU(\tb,\tc)\hatU(\tc,\ta)|\xa\ra
=\int_{-\infty}^{\infty}d^D\xc
\la\xb|\hatU(\tb,\tc)|\xc\ra\la \xc|\hatU(\tc,\ta)|\xa\ra
\label{eqgroupa}
\eeqn
so that the \tea{}s
 satify the integral relation
\beqn
(\xb\tb|\xa\ta)
=
\prod_{i=1}^D\left[ \int_{-\infty}^{\infty}
 d\xc^i\right]
(\xb\tb|\xc\tc) (\xc\tc|\xa\ta).
\label{eqgroup}
\eeqn

The matrix elements in (\ref{eq0})
have a path  integral representation
\be
\la\xb|\hatU(\tb,\ta)|\xa\ra=
\int\calD x
\,\exp\left[
\f{i}{\hbar}\calA(x)
\right]
\label{eq1a}
\ee
where
\begin{equation}
 \calA(x)   =\intab dt L(x,\dot x,t)
\label{@}\end{equation}
is the action and $L$ the Lagrangian
of the system.
The semiclassical approximation is defined rewriting
 the result of the path integration
as a product
\be
\la\xb|\hatU(\tb,\ta)|\xa\ra
\approx
\exp\left[
\f{i}{\hbar}\calAcl(\xb,\xa;\tb,\ta)
\right]\fluc(\xb,\xa;\tb,\ta)
\label{eq1}
\ee
where
 $\calAcl(\xb,\xa;\tb,\ta)$ is the associated classical action, i.e., the
action
${\cal A}[x]$ evaluated
for
the solution $\xcl(t)$ of the Euler-Lagrange classical equation of motion
which extremizes  $\calA[x]$ with fixed endpoints
at
$\xa,\ta$ and $\xb,\tb$.
In the semiclassical approximation, the factor $\fluc$
contains no $\hbar$ and contains the fluctuation determinant
arising
from the quadratic fluctuations around the classical path.
Its logarithm is the quantum-mechanical
 analog of the entropy of harmonic fluctuations
in quantum statistical mechanics.
Since the end points of the
paths are fixed, $x(\ta)=\xa$, $x(\tb)=\xb$, the boundary
 conditions for the fluctuations
$ \delta x(t)$
are of the
Dirichlet type: $ \delta x(\ta)=0$, $ \delta x(\tb)=0$.

For a
 point particle moving in a time-dependent
potential $V(x,t)$, the Lagrangian reads
\be
L=\f{M}{2}\dotx^2-V(x,t),
\ee
and the  fluctuation factor is
\beqn
&&\hspace{3cm}F(\xb,\xa;\tb,\ta)=
 \f{1}{\sqrt{2\pi i \hbar(\tb-\ta)/M}^D}
\sqrt{\f{\det_D (\p^2/\p t^2)}{\det_D [\p^2/\p t^2+{\bf V}^{(2)}(t)/M]}}^D
\label{eqgreen}
\eeqn
where ${\bf V}^{(2)}(t)$ is a $D\times D$ derivative matrix
collecting the
second
derivatives of the potential along the classical path:
\begin{equation}
  V_{ij}^{(2)}(t)=\left.\frac{\partial ^2}{\partial_i\partial_j}
V(x,t)\right|_{x=\xcl(t)},
\label{@}\end{equation}
where the indices $i,j$ denote  the vector components.
The
 fluctuation
determinants ${\det}_D$ consist of the
 product of eigenvalues of the $D\times D$ differential operator
 for  Dirichlet boundary conditions.

The fluctuation determinant is most easily evaluated
with the help of the
 Gelfand-Yaglom method
  \cite{gy}, and the result can be reexpressed
in terms of the
 Van Vleck-Pauli-Morette determinant
 \cite{vvpm}
\be
\fluc(\xb,\xa;\tb,\ta)=\f{1}{(2\pi i\hbar)^{D/2}}\left\{
\det{}_{\!D} [-\p_{\xa^i}\p_{\xb^j}\calAcl(\xb,\xa;\tb,\ta)]
\right\}^{1/2},
 \label{eqvvpm}
\ee
The minus sign inside the determinant makes
the argument of the square root  positive as long as
the classical trajectories do not reach a turning
point.
The continuation to longer intervals can always be done with the help of
Maslov indices \cite{pi}.

There exists various  ways of rewriting the right-hand side of
Eq.~(\ref{eqvvpm}).
A convenient form is obtained by
using the fact that
the momentum at the final time is given by the derivative of the classical
action:
\beqn
p_b^j&=&
\f{\p\calAcl(\xb,\xa;\tb,\ta)}{\p\xb^j},
\label{@pb}\eeqn
allowing to rewrite (\ref{eqvvpm}) as
\be
\fluc(\xb,\xa;\tb,\ta)=\f{1}{(2\pi i\hbar)^{D/2}}\left[
\det{}_{\!D}  (-\p_{\xa^i}p_b^j)
\right]^{1/2}.
 \label{eqvvpm2}
\ee

\begin{figure}[h]
\unitlength 1cm
\begin{center}
\begin{picture}(3,2)
\put(-7.5,0)
{\vbox{\begin{center}
\epsfig{file=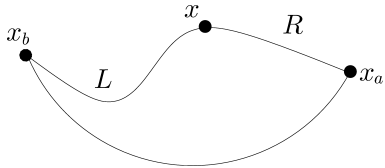,width=5cm}\end{center}}}
\end{picture}
\end{center}
\caption[]{
The upper curve shows two classical
 paths running from  $x_a$ to $x$ and from $x$ to $x_b$.
The intermediate point $x$
has to be determined by the saddle point condition.
The lower curve shows the direct classical path.
\label{figpath}}
\end{figure}

For the decomposition (\ref{eq1}) of the matrix elements of the time evolution
operator,
the group property (\ref{eqgroupa})
takes the form
\beqn
e^{\f{i}{\hbar}\calAcl(\xb,\xa;\tb,\ta)
}\fluc(\xb,\xa;\tb,\ta)&=&
\left[\prod_{i=1}^D  \int_{-\infty}^{\infty}dx^i\right]
e^{\f{i}{\hbar}\calAcl^L(\xb,x;\tb,t)
}\fluc(\xb,x;\tb,t) \times
e^{\f{i}{\hbar}\calAcl^{\!R}(x,\xa;t,\ta)
}\fluc(x,\xa;t,\ta)
\label{eqgroupsc}
\eeqn
Here we have introduced superscripts $L$ and $R$ to emphasize the left
 and right positions of the
action in the product.
The configuration of the variables is illustrated in Fig.~\ref{figpath}.

The fluctuation factor
(\ref{eqvvpm}) has the property that the semiclassical
approximation~(\ref{eq1})
satifies this equation
providing  that the intermediate $x$-integrals are evaluated in
the saddle-point approximation.
This operation is done explicitly as follows. We denote the
extremum  of the intermediate integration over $x$ by
 $\xs$, and expand the integrand  around $\xs$
 up to quadratic terms in $ \delta x\equiv x-\xs$.
Since the fluctuation factor contains no $\hbar$,
only the action in the exponent has to be expanded, and the semiclassical
approximation to
(\ref{eqgroupsc}) reads
\beqn
&&\exp\left[
\f{i}{\hbar}\calAcl(\xb,\xa;\tb,\ta)
\right]\fluc(\xb,\xa;\tb,\ta)=
\exp\left[
\f{i}{\hbar}\calAcl^{\!L}(\xb,\xcs;\tb,\tc)
+\f{i}{\hbar}\calAcl^{\!R}(\xcs,\xa;\tc,\ta)
\right]\nonumber\\
&&\hspace{2cm}\mbox{}\times
\fluc(\xb,\xcs;\tb,\tc)
\fluc(\xcs,\xa;\tc,\ta)
\int_{-\infty}^{\infty}\prod_{i=1}^D d\dx^i
\exp\left[
\f{i}{2!\hbar}\dx^i\f{\p^2}{\p\xcs^i\p\xcs^j}(\calAcl^{\!R}+\calAcl^{\!L})\dx^j
\right].
\label{eq100}
\eeqn
The saddle point condition for $\xs$ is
\be
\f{\p}{\p\xcs^i}(\calAcl^{\!R}+\calAcl^{\!L})=0.
\label{eqsaddle}
\ee
Just as in Eq.~(\ref{@pb}), the derivatives are equal to the momenta
at the intermediate time $t$,
\beqn \f{\p}{\p
\xcs^i}\calAcl^{\!R}(\xcs,\xa;\tc,\ta)&=&p_i^R(\tc),\label{rightmom}\\
\f{\p}{\p \xcs^i}\calAcl^{\!L}(\xb,\xcs;\tb,\tc)&=&-p_i^L(\tc),
\label{leftmom}\eeqn
so that the saddle point condition (\ref{eqsaddle}) implies
the equality of the intermediate momenta:
\be
p_i^L(\tc)=p_i^R(\tc).
\label{eqmomentaequality}
\ee

Our proof will be
straightforward for a
general Lagrangian which is at most quadratic in the velocitie:

\be
L(x,\dot x;t)=\f{1}{2}\dotx_ig_{ij}(x,t)\dotx_j+\dotx_i a_i(x,t)-V(x,t),\ \ \
 (g_{ij}=g_{ji}.)
\label{eqlag}
\ee
The kinetic metric $g_{ij}(x,t)$ is also known as the Hessian, whose
determinant is assumed to be nonzero to have a nondegenerate quantum system.

The canonical momenta are
\be
p_i\equiv \frac{\partial L}{\partial \dot x}=g_{ij}\dotx_j+a_i,
\label{eqmomentumdef}
\ee
and the Hamiltonian is the Legendre transform of
$L(x,\dot x;t)$:
\be
H(\tc)\equiv
p_i\dotx^i-L=\f{1}{2}[p_i(\xc,\tc)-a_i(\xc,\tc)]
g^{ij}(\xc,\tc)[p_j(\xc,\tc)-a_j(\xc,\tc)]
+V(\xc,\tc),
\label{eqh0}
\ee
where $g^{ij}(x,t)$ is the inverse matrix of
the Hessian $g_{ij}(x,t)$.

The Euler-Lagrange equations of motion following from (\ref{eqlag})
are
\be
\f{d}{dt}\f{\p L}{\p\dotx_i}-\f{\p L}{\p x_i}=0.
\label{@EL}\ee
 They are second-order
 differential equations in time.
Due to this property,
the condition
 (\ref{eqmomentaequality}) at the junction between the left and right paths
ensures that  the saddle point $\xs$ is located on the
the  classical trajectory $\xcl(t)$ running all the way
from $x_a$
to $x_b$. The upper curve of Fig.~\ref{figpath} coincide then with the lower
one.
Moreover, the  sum of the actions
is equal to the total classical action
along this combined path:
\be
\calAcl^{\!L}(\xb,\xcs;\tb,\tc)+\calAcl^{\!R}(\xcs,\xa;\tc,\ta)
=\calAcl(\xb,\xa;\tb,\ta).
\ee
Performing the Gaussian integral in Eq.~(\ref{eq100}),
we therefore find the
semiclassical consequence of the group property
(\ref{eqgroupsc}) for
of the fluctuation factor:
\be
\fluc(\xb,\xa;\tb,\ta)=
\fluc(\xb,x;\tb,t)
\fluc(x,\xa;t,\ta)
{(2\pi i\hbar)^{D/2}}{\left\{
\det{}_{\!D}  \left[\f{\p^2(\calAcl^{\!L}+\calAcl^{\!R})}{\p x^i\p x^j}\right]
\right\}^{-1/2}},
\label{eq200}
\ee
where we have omitted the tilde on top of the intermediate position
$x$ on the classical path. This equation is an algebraic
version of the eikonal equation
in Schr\"odinger theory.

It is straightforward to verify that the
Van Vleck-Pauli-Morette
fluctuation factor (\ref{eqvvpm}) satisfies (\ref{eq200}).
We shall prove this using the equivalent form (\ref{eqvvpm2}).
Inserting this into (\ref{eq200}), and using
(\ref{rightmom}) and
(\ref{leftmom}),
we have to show that
\be
\det{}_{\!D}  \left( -\left.\f{\p p_b}{\p\xa}\right|_{\xb} \right) =
\f{\det{}_{\!D}  \left(
-\left.\p p^L_b/\p\xc\right|_{\xb} \right)\det{}_{\!D}  \left( -\left.\p
p^R/\p\xa\right|_{\xc} \right) }
 {\det{}_{\!D}  \left( \left.\p p^R/\p\xc\right|_{\xa}
- \left.\p p^L/\p\xc\right|_{\xb}
  \right)},
  \label{klidentity}
\ee
where we have ignored
vector indices, for simplicity,
and
emphasized
the variable kept constant in  the partial differentiations. We also used
$p^L_b=p_b$.
The proof follows from the
chain rule for the Jacobians.
Taking the left-hand side of Eq.~(\ref{klidentity})
to the right-hand side, we must verify
that
\be
1=\f{\det{}_{\!D}  \left( -\left.\p p^R/\p\xa\right|_{\xc} \right)
\det{}_{\!D}  \left( \left.\p \xa/\p\xc\right|_{\xb} \right)}
 {\det{}_{\!D}  \left( \left.\p p^R/\p\xc\right|_{\xa}
- \left.\p p^L/\p\xc\right|_{\xb}
  \right)}.
\label{identity}
\ee
The saddle-point condition (\ref{eqmomentaequality}) is, in a more explicit
notation,
\be
p^L[\xb,\xc(\xa,\xb)]=
p^R[\xa,\xc(\xa,\xb)]. \ee
This equality allows to derive
\be
\left.
\f{\p p^L}{\p\xc}\right|_{\xb}\equiv
\left.\f{\p p^R}{\p\xc}\right|_{\xb}\equiv
\left.\f{\p
p^R[\xa,\xc(\xa,\xb)]}{\p\xc}\right|_{\xb} =
\left.\f{\p
p^R}{\p\xa}\right|_{\xc}\left.\f{\p\xa}{\p\xc}\right|_{\xb}
+\left.\f{\p p^R}{\p\xc}\right|_{\xa}. \ee
Inserting this equation into the denominator of (\ref{identity})
proves that (\ref{klidentity}) is indeed
 satisfied.

\section{Global Derivation of Fluctuation Factor}

\begin{figure}[h]
\unitlength 1cm
\begin{center}
\begin{picture}(3,2)
\put(-7.5,0)
{\vbox{\begin{center}
\epsfig{file=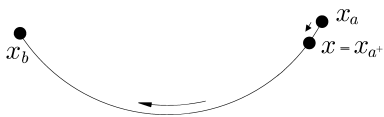,width=5cm}\end{center}}}
\end{picture}
\end{center}
\caption[]{Classical path from $\xa$ to an intermediate position $\xc=x_{a^+}$
 very
close to $\xa$, followed by
a classical path  from $x_{a^+}$  to $\xb$.
\label{path2p}}
\end{figure}

We are now prepared for the essential part of this
paper, in which
we
{\em derive\/}
the
Van Vleck-Pauli-Morette formula (\ref{eqvvpm})
from
the
semiclassical group property
(\ref{eq200})
of the fluctuation factor.
 We proceed in two steps:
 first we move the intermediate time $t$
infinitesimally close to the initial
time $t_a$. This time is called
$t_{a^+}$.
The corresponding intermediate position $x$ will then lie
at a point $x_{a^+}$ near $x_a$,
 as illustrated in Fig. \ref{path2p}.
For this configuration,
the fluctuation factor (\ref{eq200})
reads more explicitly
\be
\fluc(\xb,\xa;\tb,\ta)&=&
\fluc(\xb,\xaplus;\tb,\taplus)
\fluc(\xaplus,\xa;\taplus,\ta)
{(2\pi i\hbar)^{D/2}}\nonumber \\&&\times {\left\{
\det{}_{\!D}  \left[
\f{\p^2\calAcl^{\!L}(\xb,\xaplus;\tb,\taplus)}{\p\xc^i\p\xc^j}
+
\f{\p^2 \calAcl^{\!R}(\xaplus,\xa;\taplus,\ta)}{\p\xc^i\p\xc^j}
\right]
\right\}^{-1/2}}.
\label{eq201c}
\ee
In the limit $t_{a^+}\rightarrow t_a$
we now extract the behavior
of $\fluc(\xaplus,\xa;\taplus,\ta)$.
Intuitively,
 this should be determined
by the kinetic term of the action only, since the potential has no time
to become active.
Let us see how this comes about.
First we use the equations for the momenta
(\ref{rightmom})
and (\ref{leftmom}),
express these in terms of the derivatives of the Lagranian
with respect to the velocities
via (\ref{eqmomentumdef}),
and derive
the relations
\beqn
\f{\p^2}{\p\xc^i\p\xc^j}\calAcl^{\!R}(\xc,\xa;\tc,\ta)
&=&\f{\p}{\p\xc^i}p_j^R(\tc)=
\f{\p^2L}{\p x_i\p \dotx_j}+
\f{\p^2L}{\p \dotx_k\p \dotx_j}\f{\p}{\p \xc^i}
\dotx_k^R(\tc)
\label{@ew1}\\
\f{\p^2}{\p\xc^i\p\xc^j}\calAcl^{\!L}(\xb,\xc;\tb,\tc)
&=&-\f{\p}{\p\xc^i}p_j^L(\tc)=
-\f{\p^2L}{\p x_i\p \dotx_j}-
\f{\p^2L}{\p \dotx_k\p \dotx_j}
\f{\p}{\p \xc^i}\dotx_k^L(\tc).
\label{@ew2}\eeqn
On the right-hand side we have
taken into account that
the arguments $x$ and $\dot x$
of the Lagrangian are classical trajectories
fixed by their end points, i.e.,
$x(t)=\xcl(x,x_a;t)$ in Eq.~(\ref{@ew1}) and
$x(t)=\xcl(x_b,x;t)$ in Eq.~(\ref{@ew2}).
The derivatives with respect to the end points
produce therefore an extra term coming from the velocity dependence
of $L(x,\dot x;t)$.
Then we use the Lagrangian~(\ref{eqlag}) once more to express
\beqn
\f{\p L}{\p x_k}&=&\f{1}{2}\dotx_i[\p_kg_{ij}(x,t)]\dotx_j
+\dotx_i\p_ka_i(x,t)-\p_kV(x,t)\\
\f{\p^2 L}{\p \dotx_i\p x_j}&=&[\p_jg_{ik}(x,t)]\dotx_k+\p_ja_i(x,t)\\
\f{\p^2 L}{\p \dotx_i\p \dotx_j}&=&g_{ij}(x,t)
\eeqn
such that the brackets in Eq.~(\ref{eq201c}) lead, via
(\ref{@ew1}) and (\ref{@ew2}),
to
\beqn
&&\!\!\!\!\!\f{\p^2 \calAcl^{\!R}(\xc,\xa;\tc,\ta)}{\p\xc^i\p\xc^j}
+\f{\p^2\calAcl^{\!L}(\xb,\xc;\tb,\tc)}{\p\xc^i\p\xc^j}
\label{@35}\\
&&\hspace{0cm}=\left\{[\p_jg_{ik}(\xc,\tc)]\dotx_k(\tc)
+\p_ja_i(\xc,\tc)\right\}^R-
\left\{[\p_jg_{ik}(\xc,\tc)]\dotx_k(\tc)+\p_ja_i(\xc,\tc)\right\}^L
+g_{ik}(\xc,\tc)^R
\f{\p}{\p \xc^j}\dotx_k^R(\tc)-g_{ik}(\xc,\tc)^L
\f{\p}{\p \xc^j}\dotx_k^L(\tc).\nonumber
\eeqn
%
Then we use the fact that  $\xc(t)$ and
$\dotx(\tc)$ are continuous at the junction between the left
and right paths, such that we can collect the terms on the right-hand side to
\be
\f{\p^2 \calAcl^{\!R}(\xc,\xa;\tc,\ta)}{\p\xc^i\p\xc^j}
+\f{\p^2\calAcl^{\!L}(\xb,\xc;\tb,\tc)}{\p\xc^i\p\xc^j}=
g_{ik}(\xc,\tc)\left[
\f{\p}{\p \xc^j}\dotx_k^R(\tc)-\f{\p}{\p \xc^j}\dotx_k^L(\tc)
\right].
\label{@36}\ee
If we now take the limit $t\rightarrow \taplus$, the
contribution from the path $R$ to the derivatives
inside the brackets
becomes much larger than that of the path $L$.
Indeed, the
associated short classical path is
\be
\dotx_k^R(\taplus)\approx \f{x_k(\taplus)-x_k(\ta)}{\taplus-\ta},
\ee
implying a very large derivative
\be
\f{\p}{\p \xc^j}\dotx_k^R(\taplus)
\approx\f{\delta_{kj}}{\taplus-\ta}.
\ee
Inserting this dominant
contribution into Eq.~(\ref{@36}), this further into (\ref{eq201c}),
and factorizing out the approximately equal unknown fluctuation factors
$\fluc(\xb,\xa;\tb,\ta)\approx
\fluc(\xb,\xaplus;\tb,\taplus)
$,
we obtain
 the fluctuation factor for the
infinitesimal time interval:
\be
\fluc(\xaplus,\xa;\taplus,\ta)\approx
\f{1}{(2\pi i\hbar)^{D/2}}\left\{
\det{}_{\!D}  \left[
\f{g_{ij}(\xa,\ta)}{\taplus-\ta}
\right]
\right\}^{1/2}.
\label{eq201}
\ee
This is the well-known free-particle result, as anticipated. It will be used
twice:
first to obtain Eq.~(\ref{eq201b}), and later to fix the sign of the solution
in Eq.~(\ref{eq301bis}).

\begin{figure}[h]
\unitlength 1cm
\begin{center}
\begin{picture}(3,2)
\put(-7.5,0)
{\vbox{\begin{center}
\epsfig{file=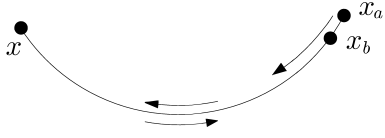,width=5cm}\end{center}}}
\end{picture}
\end{center}
\caption[Classical path from $\xa$ to $x$, followed by
a classical path backward in time from $x$ to $x_{b}$ along the same
path.]{Classical path from $\xa$ to $x$, followed by
a classical path backward in time from $x$ to $x_{b}$ along the same path.
\label{figpath2}}
\end{figure}

We turn now to the second step in the derivation.
It is based on the observation that
the group property (\ref{eq200})
 is not only valid for
causal time configurations $t_b>t>t_a$, but also for
acausal time configurations $t>t_b>t_a$.
The causality is only a property of the time evolution amplitude
(\ref{eq0}), not of the matrix elements
(\ref{eq1}), such that also (\ref{eqgroupsc})
and
(\ref{eq200}) are valid for
$t_b>t>t_a$ and
 $t>t_b>t_a$.
This means that we can also bring the time
$\tb$ close
to $\ta$, leaving $\tc$ much larger than these two adjacent times, as indicated
in
Fig.~\ref{figpath2}.
{}~\\
In this limit $\tb\rightarrow\taplus$, Eq. (\ref{eq200}) reads
\be
\fluc(\xaplus,\xa;\taplus,\ta)=
\fluc(\xaplus,\xc;\taplus,\tc)
\fluc(\xc,\xa;\tc,\ta)
{(2\pi i\hbar)^{D/2}}{\left\{
\det{}_{\!D}
\left[\f{\p^2(\calAcl^{\!L}+\calAcl^{\!R})}{\p\xc^i\p\xc^j}\right]
\right\}^{-1/2}}
\label{eq201b}
\ee
where $
\calAcl^{\!L}$ and
$\calAcl^{\!R}$ are now abbreviations for
$\calAcl^{\!L}\equiv \calAcl(\xaplus,\xc;\taplus,\tc),~\calAcl^{\!R}\equiv
\calAcl(\xc,\xa;\tc,\ta)$.
Using
(\ref{eq201}), this can be rewritten as
\be
\fluc(\xaplus,\xc;\taplus,\tc)
\fluc(\xc,\xa;\tc,\ta)
=  \f{1}{(2\pi i\hbar)^D}
\left\{
\det{}_{\!D}  \left[
\f{g_{ij}(\xa,\ta)}{\taplus-\ta}
\right]
\right\}^{1/2}
{\left\{
\det{}_{\!D}
\left[\f{\p^2(\calAcl^{\!R}+\calAcl^{\!L})}{\p\xc^i\p\xc^j}\right]
\right\}^{1/2}}.
\label{eq202}
\ee
Let us study the behavior of the brackets in the last determinant
for $t_{a^+}$ close to $\ta$.
It reads more explicitly
\be
\left.
\f{\p^2\calAcl^{\!R}(\xc,\xa;\tc,\ta)}{\p\xc^i\p\xc^j}
+\f{\p^2\calAcl^{\!L}(\xaplus,\xc;\taplus,\tc)}{\p\xc^i\p\xc^j}
\right|_{
{\taplus\approx\ta} ,
}
\label{eq203}
\ee
where
$\xaplus$ is very close to $ \xa$.
As the limit $t_{a^+}\rightarrow t$ is reached, the
 two paths coincide, and have the
 same classical action, except for a negative relative sign,
since the
 corresponding paths have the opposite direction in time.
 Thus for $t_{a^+}=t_a$, the sum in (\ref{eq203}) vanishes.
For small
 $t_{a^+}- t$,
we perform
a Taylor expansion of the second term around
the first and have, omitting the now superfluous
distinction between $L$ and $R$, and using double primes to abbreviate
the second derivatives $\partial ^2/\partial x_i\partial x_j$,
\be
\calAcl"(\xaplus,\xc;\taplus,\tc)\approx
\calAcl"(\xa,\xc;\ta,\tc)+\left[\f{\p}{\p\ta}\calAcl"(\xa,\xc;\ta,\tc)\right]
(\taplus-\ta)+
\left[\f{\p}{\p\xa^k}\calAcl"(\xa,\xc;\ta,\tc)\right](x_{a^+}^k-\xa^k) .
\ee
Inserting here
$x_{a^+}^k-\xa^k\approx (\taplus-\ta)\dotx^k(\ta)$,
we may replace (\ref{eq203})
by
\be
\left\{ \left[\f{\p}{\p\ta} \f{\p^2
\calAcl(\xa,\xc;\ta,\tc)}{\p\xc^i\p\xc^j}\right]+
\left[\f{\p}{\p\xa^k} \f{\p^2
\calAcl(\xa,\xc;\ta,\tc)}{\p\xc^i\p\xc^j}\right]\dotx^k(\ta)
\right\} (\taplus-\ta),
\label{eq204} \ee
and (\ref{eq202}) becomes
\beqn &&\fluc(\xc,\xa;\tc,\ta)\fluc(\xa,\xc;\ta,\tc)=
  \f{1}{(2\pi i\hbar)^D}{\left\{
\det{}_{\!D}  \left[ g_{ij}(\xa,\ta) \right]\right\}^{1/2} }\nonumber\\
&&\hspace{2cm}\mbox{}\times \bigglb( \det{}_{\!D}  \left\{
\left[\f{\p^2}{\p\xc^i\p\xc^j}\f{\p\calAcl(\xa,\xc;\ta,\tc)}
{\p\ta}\right] + \left[\f{\p^2}{\p\xc^i\p\xc^j} \f{\p\calAcl
(\xa,\xc;\ta,\tc)}{\p\xa^k}\right]\dotx^k(\ta) \right\}
\biggrb)^{1/2}. \label{eq205} \eeqn
Note that the derivative $\p^2/\p\xc^i\p\xc^j$ does {\em
not\/} act
 on $\dotx^k(\ta)$, so that the total
argument in the last determinant
 of (\ref{eq205}) is not the double prime of
the total derivative of the action with
respect to time $\ta$, in which case it could have been simplified to
 $(\p^2/\p\xc^i\p\xc^j)  d\calAcl(\xa,\xc;\ta,\tc)/d\ta=
(\p^2/\p\xc^i\p\xc^j)L(\xa,\dotx(\ta),\ta)$.
Since this is not the case, we can only
do a partial simplification
using the Hamilton-Jacobi equation
\be
\f{\p}{\p\ta}\calAcl(\xa,\xc;\ta,\tc)+E(\xa,\xc;\ta,\tc)=0,
\label{eqhj}
\ee
where $E(\xa,\xc;\ta,\tc)$ is the energy at the time $t_a$
for
the classical trajectory
$x(t_a)=x_{\rm cl}(x_a,x;t_a,t)$
running backwards from
$x$ at $t$ to
 $x_a$ at $t_a$.
It is the value of the
Hamiltonian $H(t_a)$ of Eq.~(\ref{eqh0})
evaluated for this trajectory.
If
$p(\ta)=p_{\rm cl}(x_a,x;t_a,t)$
denotes the associated trajectory in momentum space,
the energy is given by
\begin{eqnarray}
E(x_a,x;t_a,t)&=& H(\tc)|_{x(t)=x_{\rm cl}(x_a,x;t_a,t),p(t)=p_{\rm
cl}(x_a,x;t_a,t)},\nonumber \\
&=&\f{1}{2}[p_i(\xa,\ta)-a_i(\xa,\ta)]
g^{ij}(\xa,\ta)[p_j(\xa,\ta)-a_j(\xa,\ta)]
+V(\xa,\ta).
\label{@}\end{eqnarray}
With this,
Eq.~(\ref{eqhj}) allows to rewrite
(\ref{eq205}) as
\be
\fluc(\xc,\xa;\tc,\ta)\fluc(\xa,\xc;\ta,\tc)&=&
\f{\left\{
\det{}_{\!D}  \left[
g_{ij}(\xa,\ta)
\right]\right\}^{1/2}
}{(2\pi i\hbar)^{D}}
\nonumber \\&\times &\bigglb(
\det{}_{\!D}  \left\{
-\f{\p^2}{\p\xc^i\p\xc^j}E(x_a,x;t_a,t)
+\left[\f{\p^2}{\p\xc^i\p\xc^j}
\f{\p \calAcl(\xa,\xc;\ta,\tc)}{\p\xa^k}\right]\dotx^k(\ta)
\right\}
\biggrb)^{1/2}.
\label{eq300}
\ee
At this point we observe that for a purely harmonic
Lagrangian, which is
at most quadratic in the velocities and  positions,
the functions $g_{ij}(x,t)$ in
the general expression (\ref{eqlag}) are position-independent, the vector
potential $a_i(x,t)$ is at
most linear in $x$,
and the scalar potential $V(x,t)$
 has the general form $V=x_i\Omega_{ij}(t)x_j/2$.
Then
the classical action is at most quadratic in the end points. This
implies a vanishing second term in the brackets  of the  second
determinant in (\ref{eq300}).
Then,  we have
\be
\fluc(x,x_a;t,t_a)
^2
\mathop{=}_{\rm for~quadratic~actions}
\f{\left\{
\det{}_{\!D}  \left[
g_{ij}(\ta)
\right]\right\}^{1/2}}{(2\pi i\hbar)^{D}}
\left\{
\det{}_{\!D}  \left[
\f{\p^2}{\p\xc^i\p\xc^j}E(x_a,x;t_a,t)
\right]
\right\}^{1/2}. \ \ \
\label{eq301}
\ee
Note the sign change of the second derivative of
$E(x_a,x;t_a,t)$.
This is caused by the replacement
\begin{eqnarray}
\fluc(\xc,\xa;\tc,\ta)\fluc(\xa,\xc;\ta,\tc)\rightarrow
i^D\fluc(\xc,\xa;\tc,\ta)
^2         .
\label{@sign}\end{eqnarray}
The reason for the factor $i^D$ lies in  the Fresnel nature of the
path integral over the fluctuations.
The exponent
is the second functional
derivative of the action with
a factor $i$. Assuming stable orbits,
the factor of $i$ is
positive or negative, depending
on the
time direction of  the path.
This sign change implies that
the Fresnel integrals are related by
\begin{equation}
\fluc(\xc,\xa;\tc,\ta)\sqrt{i}^{D}=\fluc(\xa,\xc;\ta,\tc)/\sqrt{i}^{D}.
\label{@}\end{equation}
We can now take  the square root of (\ref{eq301}) and obtain
\be
\fluc(\xb,\xa;\tb,\ta)
\mathop{=}_{\rm for~quadratic~actions}
\f{\left\{
\det{}_{\!D}  \left[
g_{ij}(\ta)
\right]\right\}^{1/4}}{(2\pi i\hbar)^{D/2}}
\left\{
\det{}_{\!D}  \left[
\f{\p^2}{\p\xb^i\p\xb^j}E(x_a,x_b;t_a,t_b)
\right]
\right\}^{1/4}, \ \ \
\label{eq301bis}
\ee

The sign of this square root
is fixed by
the fact that
in the limit of short intervals $\tb-\ta$, the fluctuation factor has to
reduce to
 the free-particle result~(\ref{eq201}).

\comment{The first determinant in~(\ref{eq301bis}) depends only on the
coefficient
 $g_{ij}(t_a)$ of the kinetic  energy of the Hamiltonian.\mn{need to discuss}
But also the second term can be evaluated using only this information.
 This comes from the fact that we need to evaluate the second
derivative with respect to $\xc$ of the Hamiltonian evaluated at $\ta$, leading
to $\p V(\xa,\ta)/\p\xc=0$.
However, for pratical purpose, it may be easier to
work with the Hamiltonian expression rather than with the kinetic term alone:
the total energy is simpler to consider than its partial contributions.}

Note that in the semiclassical limit the fluctuations are
always harmonic. For a vanishing vector potential $a_i(x,t)$ in
Eq.~(\ref{eqlag}),
these would be driven
by a time-dependent frequency matrix
\begin{equation}
 \Omega _{ij}(t)=g^{ik}(x_{\rm cl}(t),t) \p_k\p_jV(x_{\rm cl}(t),t).
\label{@freq}
\end{equation}
 This harmonic property does not, however, allow us to use
formula (\ref{eq301bis})
for the fluctuation factor, since the frequency matrix depends
on the end
 points via
the classical  solution
$\xcl(t)$ of the equations of motion,
 so that
the
 full
formula (\ref{eq300}) must be used, which we now investigate in detail.

We must evaluate
\be
-\f{\p^2}{\p\xc^i\p\xc^j}E(x_a,x;t_a,t)
+\left[\f{\p^2}{\p\xc^i\p\xc^j}
\f{\p \calAcl(\xa,\xc;\ta,\tc)}{\p\xa^k}\right]\dotx^k(\ta)
\label{eq310}
\ee
with $\p\calAcl(\xa,\xc;\ta,\tc)/\p\xa^k=p_k(\ta)$.
Since $\p\xa^k/\p\xc^j=0$, expression (\ref{eq310}) can be rewritten as
\be
-\f{1}{2}g_{kl}(\xa,\ta)\f{\p^2}{\p\xc^i\p\xc^j}\dotx_k(\ta)\dotx_l(\ta)
+\dotx_k(\ta)g_{kl}(\xa,\ta)\f{\p^2}{\p\xc^i\p\xc^j}\dotx_l(\ta).
\ee
Using the symmetry $g_{kl}(\xa,\ta)=g_{lk}(\xa,\ta)$, this becomes
\be
-g_{kl}(\xa,\ta)\f{\p}{\p\xc^i}
\dotx_k(\ta)\f{\p}{\p\xc^i}\dotx_l(\ta).
\ee
Now the determinant of a product of matrices  factorizes
into a product of determinants,
and Eq.~(\ref{eq300})  becomes
\be
\fluc(\xc,\xa;\tc,\ta)
^2=
\f{
\det{}_{\!D}  \left[
g_{ij}(\xa,\ta)
\right]}
{(2\pi i\hbar)^{D}}
\det{}_{\!D}  \left[
\f{\p}{\p\xc^i}\dotx_j(\ta)
\right]
,
\label{eq311}
\ee
where the phase factor (\ref{@sign})
has
been taken into account.
Using the relation
\be
\f{\p^2}{\p\xa^i\p\xc^j}\calAcl(\xc,\xa;\tc,\ta)=
-\f{\p}{\p\xc^j}p_i(\ta)=-g_{ik}(\xa,\ta)\f{\p}{\p\xc^j}\dotx_k(\ta)
\ee
we can finally rewrite (\ref{eq311}) as
\be
\fluc(\xc,\xa;\tc,\ta)
^2=
\f{1}{(2\pi i\hbar)^{D}}
\det{}_{\!D}  \left[
-\f{\p^2}{\p\xa^i\p\xc^j}\calAcl(\xc,\xa;\tc,\ta)
\right]
\ee
from which it is straightforward to obtain
\be
\fluc(\xb,\xa;\tb,\ta)=\f{1}{(2\pi i\hbar)^{D/2}}
\left\{
\det{}_{\!D}  \left[-
\f{\p^2}{\p\xa^i\p\xb^j}\calAcl(\xb,\xa,\tb,\ta)
\right]
\right\}^{1/2}.
\label{finalsol}
\ee
This is the Van Vleck-Pauli-Morette formula (\ref{eqvvpm}).

Our derivation has ignored
zero modes in the intermediate integration, which
may be  treated in the standard way \cite{pi}.

\section{Applications}
\label{sectionapplication}

Here we shall apply our formula
to three systems of point particles:\\[1em]
\phantom{xxxxx}A.  the
 free point particle with a mass matrix,
with a Lagrangian
\be
L=\f{1}{2}\dotx_iM_{ij}\dotx_j,
\label{free}
\ee
\phantom{xxxxx}B. the harmonic oscillator  space with a
time-dependent
frequency matrix
$\omega_{ij}(t)$, a mass matrix and a Lagrangian
\be
L=\f{1}{2}\left\{ \dotx_iM_{ij}\dotx_j-x_i[M\omega^2(t)]_{ij}x_j\right\} ,
\label{ho}
\ee
\phantom{xxxxx}C. an ordinary particle in a constant magnetic field
perpendicular to the plan
spanned by  $x_1$ and $x_2$,
 with a \phantom{xxxxx3. }Lagrangian
\be
L=L=\f{1}{2}M\sum_{i=1}^D\dotx_i^2-\f{e}{c}B\dotx_2x_1.
\label{mag}
\ee
In each cases, the boundary conditions are $x(\ta)=\xa,x(\tb)=\xb$.
\subsection{Free Particle}

The free particle case is particularily simple. The Lagrangian (\ref{free})
implies the equations of motion
\be
M_{ij}\f{d^2}{dt^2}x_j=0.
\ee
Since
the
matrix
${\mathbf M}$ is symmetic, it can be diagonalized by a similarity
transformation
with an orthogonal  matrix
${\mathbf S}$=
${\mathbf S}^{-T}$. Let
${\mathbf M}^d={\mathbf S}^{-1}{\mathbf M}{\mathbf S}$
be the resulting
 diagonal mass matrix.
The normal modes of the motion are $y(t)={\mathbf S}^{-1}x(t)$. The latter
satisfy
\be
\f{d^2}{dt^2}y_j=0
\ee
and have the time dependence
\be
y_i(t)=\f{1}{\tb-\ta}\left[
y_a^i(\tb-t)+y_b^i(t-\ta)
\right].
\ee
The associated classical Hamiltonian is
\be
H=\f{1}{2}\dotx_iM_{ij}\dotx_j=\f{1}{2}\doty_iM^d_{ij}\doty_j,
\ee
and the trajectories have the energy
\be
E(x_a,x_b;t_a,t_b )
=\f{1}{2}\sum_{k=1}^D\f{(y_b^k-y_a^k)M^d_{kk}(y_b^k-y_a^k)}{(\tb-\ta)^2},
\ee
Using the relation
\be
\f{\p}{\p\xb^i}=\f{\p y_b^k}{\p\xb^i}\f{\p}{\p y_b^k}=S^{-1}_{ki}\f{\p}{\p
  y_b^k},
\ee
we deduce
\be
\f{\p^2}{\p\xb^i\p\xb^j}E(x_a,x_b;t_a,t_b )=\f{M_{ij}}{(\tb-\ta)^2}.
\ee
Inserting this into Eq.~(\ref{eq301bis}), we obtain the well-known
fluctuation factor
\be
\fluc(\xb,\xa;\tb,\ta)=\f{\left(\det{}_{\!D}   {\mathbf M}\right)^{1/2}}
{\left[2\pi i\hbar(\tb-\ta)\right]^{D/2}}.
\ee

\subsection{Harmonic Oscillator with Time-Dependent Frequency}

The case of the harmonic oscillator
with a time-dependent frequency is slightly more involved. One cannot solve
the equations of motion to get the solution in a closed form. We will however
give a formal solution, showing how the well-known result can be recovered
when the frequency is time independent. The equations
of
motion associated with (\ref{ho}) are
\be
\f{d^2}{dt^2}x_i+\omega^2_{ij}(t)x_j=0.
\label{timeho}
\ee
With the help of two matrices ${\mathbf A}$ and ${\mathbf B}$, the solution can
be decomposed as
$x={\mathbf A}\xa+{\mathbf B}\xb$. Since $\xa$ and $\xb$ are independent, each
of the matrices
satisfies a same equation as~(\ref{timeho}):
\beqn
\f{d^2}{dt^2}{\mathbf A}+{\omegabi}^2(t){\mathbf A}&=&0,\\
\f{d^2}{dt^2}{\mathbf B}+{\omegabi}^2(t){\mathbf B}&=&0\label{eqB}.
\eeqn
The  boundary conditions are
\beqn
A_{ij}(\ta)&=&\delta_{ij}, A_{ij}(\tb)=0,\\
B_{ij}(\ta)&=&0, B_{ij}(\tb)=\delta_{ij}.\label{dirichB}
\eeqn
Formula~(\ref{eq301bis}) contains a double derivative with
respect to the end  point $\xb$. For this reason, it
will depend only
on the part of the solution
with the matrix ${\mathbf B}$.
This simplifies the evaluation of the Hamiltonian
 and, taking derivatives
with respect to the end point $\xb$, we have
\be
\f{\p^2}{\p\xb^i\p\xb^j}E(x_a,x_b;t_a,t_b)= \left\{
\left[\f{d}{dt}{\mathbf B}(t)\right]_{t=\ta}{\mathbf M}
\left[\f{d}{dt}{\mathbf B}(t)\right]_{t=\ta}
\right\}_{ij}=\left[{\mathbf \dotB}(\ta)
  {\mathbf M}{\mathbf \dotB}(\ta)\right]_{ij}
\ee
where the last equality defines ${\mathbf \dotB}$ as the time derivative of the
matrix
${\mathbf B}$.
Using this relation in Eq.~(\ref{eq301bis}), the fluctuation factor is
given by
\be
\fluc(\xb,\xa;\tb,\ta)=\f{\left(\det{}_{\!D}   {\mathbf M}\right)^{1/2}}
{\left[2\pi i\hbar(\tb-\ta)\right]^{D/2}}
\left[\det{}_{\!D}  {\mathbf \dotB}(\ta)\right]^{1/2},
\label{flucmitB}
\ee
which requires to solve Eq.~(\ref{eqB}) with the associated boundary
conditions~(\ref{dirichB}). A formal solution can be obtained in the following
way. Integrating twice~(\ref{eqB}), using~(\ref{dirichB}), leads to

\be
{\mathbf B}(t)=\int_{\ta}^tds
{\mathbf \dotB}(\ta)-\int_{\ta}^tds\int_{\ta}^s
{\omegabi^2}(y){\mathbf B}(y)dy
\ee
which can be iterated to lead to a Neumann series
\be
{\mathbf B}(t)=\int_{\ta}^tds
\left\{
\one-\int_{\ta}^s{\omegabi}^2(y)dy\int_{\ta}^yds'
+\int_{\ta}^s{\omegabi}^2(y)dy\int_{\ta}^yds'
\int_{\ta}^{s'}{\omegabi}^2(y')dy'\int_{\ta}^{y'}ds''-\cdots
\right\}{\mathbf \dotB}(\ta).
\label{vnseries}
\ee

Using a first order differential formalism, this expansion can be given a
compact notation. This comes from the fact that the solution of~(\ref{eqB})
can be written as
\beqn
\left(\begin{array}{cc}
{\mathbf B}(t)&{\mathbf 0}\\
{\mathbf 0}&{\mathbf \dotB}(t)
\end{array}\right)&=&
T\left\{
\cosh\left[\int_{\ta}^t\left(\begin{array}{cc}
{\mathbf 0}&{\mathbf 1}\\
-{\omegabi}^2(s)&{\mathbf 0}
\end{array}\right)ds\right]
\right\}
\left(\begin{array}{cc}
{\mathbf B}(\ta)&{\mathbf 0}\\
{\mathbf 0}&{\mathbf \dotB}(\ta)
\end{array}\right)
+
T\left\{
\sinh\left[\int_{\ta}^t\left(\begin{array}{cc}
{\mathbf 0}&{\mathbf 1}\\
-{\omegabi}^2(s)&{\mathbf 0}
\end{array}\right)ds\right]
\right\}
\left(\begin{array}{cc}
{\mathbf 0}&{\mathbf B}(\ta)\\
{\mathbf \dotB}(\ta)&{\mathbf 0}
\end{array}\right),
\nonumber \\&&
\eeqn
where the hyperbolic functions are defined through their Taylor expansion and
where the symbol $T$ implies a time ordering operation.
Using the boundary conditions, we end up with
\be
\left(\begin{array}{cc}
{\mathbf B}(t)&{\mathbf 0}\\
{\mathbf 0}&{\mathbf 0}
\end{array}\right)=
T\left\{
\sinh\left[\int_{\ta}^t\left(\begin{array}{cc}
{\mathbf 0}&{\mathbf 1}\\
-{\omegabi}^2(s)&{\mathbf 0}
\end{array}\right)ds\right]
\right\}
\left(\begin{array}{cc}
{\mathbf 0}&{\mathbf 0}\\
{\mathbf \dotB}(\ta)&{\mathbf 0}
\end{array}\right),
\ee
where we have multiplied from the right by an appropriate matrix
 in order to single out the upper left component.
 We can extract ${\mathbf \dotB}(\ta)$ from this relation using the boundary
 condition ${\mathbf B}(\tb)=\one$. This gives
\be
{\mathbf \dotB}(\ta)=
\Bigglb(T\left\{
\sinh\left[\int_{\ta}^{\tb}\left(\begin{array}{cc}
{\mathbf 0}&{\mathbf 1}\\
-{\omegabi}^2(s)&{\mathbf 0}
\end{array}\right)ds\right]
\right\}_{12}\Biggrb)^{-1},
\label{formalseries}
\ee
which is indeed equivalent to iterative solution (\ref{vnseries}).
Using this in (\ref{flucmitB}) provides then us with a formal solution for the
fluctuation factor.

%

The case of a time independent frequency matrix is obtained directly from the
formal series~(\ref{formalseries}): the time ordering operator disappears, the
time integration is trivial and the series easily evaluated:
\be
T\left\{
\sinh\left[\int_{\ta}^{\tb}\left(\begin{array}{cc}
{\mathbf 0}&{\mathbf 1}\\
-{\omegabi}^2(s)&{\mathbf 0}
\end{array}\right)ds\right]
\right\}
\mathop{=}_{{\omegabi} \rm
  =const.}
\sin\left[{\omegabi}(\tb-\ta)\right]
\left(\begin{array}{cc}
{\mathbf 0}&{\omegabi}^{-1}\\
-{\omegabi}&{\mathbf 0}
\end{array}\right).
\ee
As stipulated in (\ref{formalseries}), we need only the upper-right component.
We then end up with
\be
\fluc(\xb,\xa;\tb,\ta)\mathop{=}_{{\omegabi} \rm
  =const.}\f{\left(\det{}_{\!D}   {\mathbf M}\right)^{1/2}}
{\left[2\pi i\hbar(\tb-\ta)\right]^{D/2}}
\left\{\f{\det{}_{\!D}  {\omegabi}}{
\det{}_{\!D}  \sin[{\omegabi}(\tb-\ta)]}\right\}^{1/2}.
\ee
Denoting by $\omega_i^2$ (no summation over $i$) the normal modes, this
equation can also be written as
\be
\fluc(\xb,\xa;\tb,\ta)\mathop{=}_{{\omegabi} \rm
  =const.}\f{\left(\det{}_{\!D}   {\mathbf M}\right)^{1/2}}
{\left[2\pi i\hbar(\tb-\ta)\right]^{D/2}}
\prod_{i=1}^D
\left\{\f{\omega_i}{\sin[\omega_i(\tb-\ta)]}\right\}^{1/2}.
\ee

\subsection{Particle in Constant Magnetic Field}

Here the calculation  looks somewhat more complicated, although it is still
trivial.
The equations of motion associated with the
Lagrangian~(\ref{mag}) are
\beqn
\f{d^2}{dt^2}x_1-\omega\f{d}{dt}x_2&=&0,\label{magmotion1}\\
\f{d^2}{dt^2}x_2+\omega\f{d}{dt}x_1&=&0,\label{magmotion2}\\
\f{d^2}{dt^2}x_j&=&0,~~~~~j=3,\dots,D.
\eeqn
The index  $j$ will be
limited to $j=3,\cdots,D$
throughout in this section.
The frequency $\omega$ is the Larmor
frequency
$ \omega =
eB/(cM)$, where $e$ is the electron charge and $c$ the speed of light.
The classical trajectories are
\beqn
x_1&=&\f{1}{\sin\omega(\tb-\ta)}\left[
(\xb^1-x_0^1)\sin\omega(t-\ta)+(\xa^1-x_0^1)\sin\omega(\tb-t)\right]+x_0^1,\\
x_2&=&\f{1}{\sin\omega(\tb-\ta)}\left[
(\xb^2-x_0^2)\sin\omega(t-\ta)+(\xa^2-x_0^2)\sin\omega(\tb-t)\right]+x_0^2,\\
x_j&=&\f{1}{\tb-\ta}\left[
\xa^j(\tb-t)+\xb^j(t-\ta)
\right],
\eeqn
where $x_0^1$ and $x_0^2$ are determined from (\ref{magmotion1})
and (\ref{magmotion2}) as \cite{pi}:
\beqn
x_0^1&=&\f{1}{2}\left[
(\xb^1+\xa^1)+(\xb^2-\xa^2)\cot\f{\omega(\tb-\ta)}{2}
\right],\\
x_0^2&=&\f{1}{2}\left[
(\xb^2+\xa^2)-(\xb^1-\xa^1)\cot\f{\omega(\tb-\ta)}{2}
\right].
\eeqn
For the
classical Hamiltonian we
 obtain the only
non-vanishing contributions (no summation over $j$)
\beqn
\f{\p^2}{\p\xb^1\p\xb^1}E(x_a,x_b;t_a,t_b)
&=&\f{\p^2}{\p\xb^2\p\xb^2}E(x_a,x_b;t_a,t_b)=
\f{M\omega^2}{4\sin^2\left[\omega(\tb-\ta)/2 \right]},\\
\f{\p^2}{\p\xb^j\p\xb^j}E(x_a,x_b;t_a,t_b)&=&\f{M}{(\tb-\ta)^2},
\eeqn
from which it is trivial to find
\be
\det{}_{\!D}  \f{\p^2}{\p\xb^i\p\xb^k}E(x_a,x_b;t_a,t_b)
=\left[
\f{M}{(\tb-\ta)^2}
\right]^{D-2}
\left\{
\f{M\omega^2}{4\sin^2\left[\omega(\tb-\ta)/2 \right]}
\right\}^2,
\ee
such that Eq.~(\ref{eq301bis}) yields  \cite{pi}
\be
\fluc(\xb,\xa;\tb,\ta)=\sqrt{\f{M}{2\pi i \hbar(\tb-\ta)}}^{D}
\f{\omega(\tb-\ta)/2}{\sin\left[\omega(\tb-\ta)/2\right]}.
\ee

\subsection{Particle in Arbitrary One-Dimensional Potential}

For a particle moving in an arbitrary time-dependent
potential in one dimension, it is possible to
construct an explicit solution to the general
relation~(\ref{eq200}).
The matrix in the  determinant on the right-hand side is
\be
\f{\p}{\p\xc}\dotx^R(\tc)-\f{\p}{\p\xc}\dotx^L(\tc).
\ee
Using the equation of motion
\be
\f{d^2}{dt^2}x+\p_xV(x,t)=0
\ee
and taking the derivative with respect to time
we obtain
\beqn
\f{d^2}{dt^2}\dotx+V^{''}(x,t)\dotx&=&0,
\label{speed1}
\eeqn
and a similar equation
for  a derivative with respect
to any  other parameter $\lambda$:
\beqn
\f{d^2}{dt^2}\p_{\lambda}x+V^{''}(x,t)\p_{\lambda}x&=&0.
\eeqn
Hence $\p_{\lambda}x$ can be expressed as
a linear combination of two fundamantal solutions of~(\ref{speed1}). One of
them is the time derivative of the classical trajectory,
$\p_{\lambda}x^{(1)}=\dotx$.
The other
can be obtain from the D'Alembert construction \cite{pi} and is
$\p_{\lambda}x^{(2)}=\dotx\int^{t}dt/\dotx^2$.
Combining these into solutions satisfying the boundary conditions
$\p_{\xc}x(\ta)=0$, $\p_{\xc}x(\tb)=0$ and $\p_{\xc}x(\tc)=1$, we then
 obtain
\beqn
\p_{\xc}x^R&=&\f{\dotx^R}{\dotx(\tc)}
\f{\int_{\ta}^tdt/\dotx^2}{\int_{\ta}^{\tc}dt/\dotx^2},\\
\p_{\xc}x^L&=&\f{\dotx^L}{\dotx(\tc)}
\f{\int_{t}^{\tb}dt/\dotx^2}{\int_{\tc}^{\tb}dt/\dotx^2},
\eeqn
from which we deduce
\be
\partial _{x}\dotx^R(\tc)-\partial _{x}\dotx^L(\tc)=\f{1}{\dotx^2(\tc)}
\f{\int_{\ta}^{\tb}dt/\dotx^2}
{\left(
\int_{\ta}^{\tc}dt/\dotx^2
\right)
\left(
\int_{\tc}^{\tb}dt/\dotx^2
\right)
}.
\ee
Inserting this result in (\ref{eq200}), we have in the limit
 $\tc\rightarrow\taplus$
\be
\left(
\dotx^2(\ta)\int_{\ta}^{\taplus}\frac{dt}{\dotx^2}
\right)^{-1/2}=\sqrt{2\pi i \hbar}\fluc(\xaplus,\xa;\taplus,\ta).
\label{shorttime}
\ee
Taking the limit
$\tb\rightarrow\ta$ at an
arbitrary $t$ in (\ref{eq200}), and inserting (\ref{shorttime}) , we obtain
\be
\left[
\dotx^2(\ta)\dotx^2(\tc)\left(
\int_{\ta}^{\tc}\frac{dt}{\dotx^2}
\right)\left(
\int_{\tc}^{\ta}\frac{dt}{\dotx^2}
\right)
\right]^{-1/2}=(2\pi i \hbar)
\fluc(\xa,\xc;\ta,\tc)\fluc(\xc,\xa;\tc,\ta).
\ee
and thus a fluctuation factor
\be
\fluc(\xb,\xa;\tb,\ta)=
\f{1}{\sqrt{2\pi i \hbar}}\left[
\dotx(\ta)\dotx(\tb)\int_{\ta}^{\tb}\frac{dt}{\dotx^2},
\right]^{-1/2}.
\ee
which agrees, of course, with formula
(\ref{eqvvpm}).

\section{Conclusion}

We have shown that the Van Vleck-Pauli-Morette
determinant in the fluctuation factor (\ref{eqvvpm})
can be
obtained directly from the group property of the
time evolution
operator~(\ref{eq0}) and the semiclassical expansion~(\ref{eq1}).
In addition, we have derived a formula which allows us
to find the fluctuation factor
form
the classical Hamiltonian function
if the Lagrangian is
at harmonic
 velocities and
 positions.

\begin{acknowledgments}
We thank A. Pelster for discussions.
\end{acknowledgments}

\end{document}